\documentclass[aps,preprint,groupedaddress,showpacs]{revtex4-1}
\usepackage{amssymb}
\usepackage[breaklinks,colorlinks,citecolor=green,urlcolor=blue,linkcolor=red]{hyperref}

\begin{document}


\title{The pollution to the $K\pi$-puzzle from the isospin-breaking $\pi^0\!\!-\!\eta\!-\!\eta'$ mixing effect}


\author{Zhen-Hua Zhang}
\affiliation{College of Nuclear Science and Technology, University of South China, Hengyang, 421001, Hunan, China}


\date{\today}

\begin{abstract}
The influence of the isospin-breaking $\pi^0\!\!-\!\eta\!-\!\eta'$ mixing effect on the $CP$-asymmetries of $B\to K\pi$ processes is examined for the first time.
It is found that this mixing effect brings a large uncertainty both to the $CP$-asymmetry sum rule of $B\to K\pi$ processes and the $CP$-asymmetry difference of the $B^+\to K^+\pi^0$ and $B^0\to K^+\pi^-$, obscuring the significance of the $K\pi$-puzzle.
This uncertainty can be so large that it is even possible to explain the $K\pi$-puzzle by the $\pi^0\!\!-\!\eta\!-\!\eta'$ mixing effect {\it alone}.
\end{abstract}


\maketitle


\section{Brief introduction to the $K \pi$-puzzle}

It is believed that the decay processes $B\to K\pi$ are good probes for new physics (NP) beyond the Standard Model (SM), as the tree-level amplitudes are suppressed, making them more sensitive to the potentially NP contributions.
Based on the isospin consideration, the amplitudes of the aforementioned weak decays $B\to K\pi$ are related through \cite{Nir:1991cu,Lipkin:1991st,Gronau:1991dq}
\begin{equation}
  \mathcal{A}_{B^+\to K^0\pi^+}+\sqrt{2}\mathcal{A}_{B^+\to K^+\pi^0}=\mathcal{A}_{{B^0}\to K^+\pi^-}+\sqrt{2}\mathcal{A}_{{B^0}\to K^0\pi^0},
\end{equation}
from which a sum rule between $CP$ asymmetries of $B\to K\pi$ processes can be deduced \cite{Gronau:2005kz}:
\begin{equation}\label{SumRule}
A_{CP}^{K^+\pi^-}+\frac{\mathcal{B}(K^0\pi^+)}{\mathcal{B}(K^+\pi^-)}\frac{\tau_0}{\tau_+}A_{CP}^{K^0\pi^+} =\frac{2\mathcal{B}(K^+\pi^0)}{\mathcal{B}(K^+\pi^-)}\frac{\tau_0}{\tau_+}A_{CP}^{K^+\pi^0}+\frac{2\mathcal{B}(K^0\pi^0)}{\mathcal{B}(K^+\pi^-)}\frac{\tau_0}{\tau_+}A_{CP}^{K^0\pi^0}.
\end{equation}
This $CP$-asymmetry sum rule can be further simplified into a more crude relation between the $CP$ asymmetries of $B^+\to K^+\pi^0$ and $B^0\to K^+\pi^-$ \cite{Gronau:1998ep}:
\begin{equation}\label{RoughRelation}
A_{CP}^{K^+\pi^0}-A_{CP}^{K^+\pi^-} \approx 0,
\end{equation}
which is clearly in contradiction to the latest world average of the $CP$-asymmetry difference between these two aforementioned processes \cite{Belle:2012dmz,BaBar:2012fgk,CDF:2014pzb,LHCb:2018pff,HFLAV:2019otj,LHCb:2020dpr},
\begin{equation}
\Delta A_{CP}(K\pi)\equiv A_{CP}^{K^+\pi^0}-A_{CP}^{K^+\pi^-}=0.115\pm0.014.
\end{equation}
This is basically a short version of the long-standing ``$K\pi$-puzzle''.

The $K\pi$-puzzle has gained constant theoretical attentions, especially those works on the exploration of the possibility of NP \cite{Buras:2003yc,Buras:2003dj,Buras:2004ub,Barger:2004hn,Baek:2004rp,Chang:2009wt,Fleischer:2017vrb,Beaudry:2017gtw,Bhattacharya:2021shk}.
Meanwhile, there are also works attempting to understand it within SM \cite{Bauer:2005kd,Cheng:2009cn,Cheng:2014rfa,Zhou:2016jkv,Kundu:2021emt}.
In this paper, we will try to examine the role of the isospin breaking effect --the $\pi^0\!\!-\!\eta\!-\!\eta'$ mixing effect-- in the $K\pi$-puzzle.

\section{the pollution of the $\pi^0\!\!-\!\eta\!-\!\eta'$ mixing effects to the $K\pi$-puzzle}

The basic idea is very simple.
Since there are $\pi^0$s in the final states of these $B\to K\pi$ decay processes, the isospin-breaking $\pi^0\!\!-\!\eta\!-\!\eta'$ mixing effect \cite{Gross:1979ur,Leutwyler:1996np,Feldmann:1998sh,Escribano:2020jdy,Kordov:2021eqx} takes place.
Although this effect seems to be negligible at first sight as it is small \cite{Gronau:2006eb} --perhaps that is why the $\pi^0\!\!-\!\eta\!-\!\eta'$ mixing effect has never been put on the table dealing with the $K\pi$-puzzle-- this kind of isospin-breaking effects could potentially affect $CP$ asymmetries more badly than expected \cite{Zhang:2020rol}.

Within the context of the $\pi^0\!\!-\!\eta\!-\!\eta'$ mixing effect, the $\pi^0$ meson can be expressed as an admixture of the isospin eigenstate $\pi_3$ and the mass eigenstates $\eta$ and $\eta^\prime$, $|\pi^0\rangle=|\pi_3\rangle+\epsilon|\eta\rangle+\epsilon^\prime|\eta^\prime\rangle$, where $\epsilon$ and $\epsilon^\prime$ are small parameters which account for the mixing between $\pi^0$ and $\eta$ and $\eta^\prime$, respectively \cite{Leutwyler:1996np,Feldmann:1998sh}.
The $CP$ asymmetry parameters of $B^{+}\to  K^{+}\pi^0$ and $B^{0}\to  K^{0}\pi^0$ acquire corrections because of the $\pi^0\!\!-\!\eta\!-\!\eta'$ mixing.
For example, up to $\mathcal{O}({\epsilon^{(')}})$, the $CP$ asymmetry of $B^+\to K^+ \pi^0 $ is modified into
\footnote{An extra factor $\left(1-{A_{CP}^{K^+\pi_3 }}^2\right)$ in front of $\Delta_{\text{IB}}^{K^+\pi^0 }$ is omitted as it is numerically very close to one.}
\begin{equation}\label{ACPcorrection}
  A_{CP}^{ K^+ \pi^0}= A_{CP}^{ K^+\pi_3}+\Delta_{\text{IB}}^{K^+\pi^0 },
\end{equation}
where $A_{CP}^{K^+\pi_3 }$ is in fact the conventionally defined $CP$ asymmetry of $B^+\to K^+\pi^0$ without the $\pi^0\!\!-\!\eta\!-\!\eta'$ mixing effect, which takes the form $A_{CP}^{K^+\pi_3 }\equiv \frac{\left|\mathcal{A}_{B^-\to K^-\pi_3 }\right|^2-\left|\mathcal{A}_{B^+\to  K^+\pi_3}\right|^2}{\left|\mathcal{A}_{B^-\to  K^-\pi_3}\right|^2+\left|\mathcal{A}_{B^+\to  K^+\pi_3}\right|^2}$,
and $\Delta_{\text{IB}}^{ K^+\pi^0}$ represents the correction of the $\pi^0\!\!-\!\eta\!-\!\eta'$ mixing effect to the $CP$ asymmetry of $B^+\to K^+ \pi^0$, which, after some algebra, can be expressed as
\begin{equation}\label{DeltaIBKpm}
  \Delta_{\text{IB}}^{K^+\pi^0 }=\Re\left[\left(\frac{\mathcal{A}_{B^-\to K^-\eta}}{\mathcal{A}_{B^-\to K^-\pi^0 }}-\frac{\mathcal{A}_{B^+\to K^+\eta}}{\mathcal{A}_{B^+\to K^+\pi^0}}\right)e^{i\theta}\epsilon+\left(\frac{\mathcal{A}_{B^-\to K^-\eta^{\prime}}}{\mathcal{A}_{B^-\to K^-\pi^0}}-\frac{\mathcal{A}_{B^+\to K^+\eta^{\prime}}}{\mathcal{A}_{B^+\to K^+\pi^0}}\right)e^{i\theta'}\epsilon'\right].
\end{equation}
Note that two strong phases between the amplitudes of $B^\pm\to K^\pm\eta^{(\prime)}$ and $B^\pm\to K^\pm\pi^0$, $\theta$ and $\theta'$, are presented explicitly.
From Eq. (\ref{ACPcorrection}) and (\ref{DeltaIBKpm}) one can roughly see that the $\pi^0\!\!-\!\eta\!-\!\eta'$ mixing correction term can not be simply neglected, provided that the $CP$ asymmetry parameter $A_{CP}^{ K^+ \pi^0}$ is about the same order with the mixing parameters $\epsilon$ and/or $\epsilon'$.
Similarly, there is also a correction term to the $CP$ asymmetry of $B^0\to K^0\pi^0$, which reads
\begin{equation}\label{DeltaIBK0}
  \Delta_{\text{IB}}^{ K^0\pi^0}=\Re\left[\left(\frac{\mathcal{A}_{\overline{B^0}\to \overline{K^0}\eta}}{\mathcal{A}_{\overline{B^0}\to \overline{K^0}\pi^0}}-\frac{\mathcal{A}_{B^0\to K^0\eta}}{\mathcal{A}_{B^0\to K^0\pi^0}}\right)e^{i\tilde{\theta}}\epsilon+\left(\frac{\mathcal{A}_{\overline{B^0}\to \overline{K^0}\eta^{\prime}}}{\mathcal{A}_{\overline{B^0}\to \overline{K^0}\pi^0}}-\frac{\mathcal{A}_{B^0\to K^0\eta^{\prime}}}{\mathcal{A}_{B^0\to K^0\pi^0}}\right)e^{i\tilde{\theta}'}\epsilon'\right].
\end{equation}

Since Eqs. (\ref{SumRule}) and (\ref{RoughRelation}) are obtained under the ignorance of the the $\pi^0\!\!-\!\eta\!-\!\eta'$ mixing effect, the $\pi^0$s in these two equations are in fact the isospin eigenstates $\pi_3$s.
Consequently, Eqs. (\ref{SumRule}) and (\ref{RoughRelation}) should be rewritten as
\begin{equation}\label{SumRuleNonPhysical}
A_{CP}^{K^+\pi^-}+\frac{\mathcal{B}(K^0\pi^+)}{\mathcal{B}(K^+\pi^-)}\frac{\tau_0}{\tau_+}A_{CP}^{K^0\pi^+} =\frac{2\mathcal{B}(K^+\pi_3)}{\mathcal{B}(K^+\pi^-)}\frac{\tau_0}{\tau_+}A_{CP}^{K^+\pi_3} +\frac{2\mathcal{B}(K^0\pi_3)}{\mathcal{B}(K^+\pi^-)}\frac{\tau_0}{\tau_+}A_{CP}^{K^0\pi_3},
\end{equation}
and
\begin{equation}\label{RoughRelationIB}
A_{CP}^{K^+\pi_3}-A_{CP}^{K^+\pi^-}\approx 0,
\end{equation}
respectively.
It should be pointed out that all the branching ratios and $CP$ asymmetries containing $\pi_3$ in the final state are strictly speaking not physical observables.
This however, can be fixed with the aid of Eq. (\ref{ACPcorrection}) by rewriting Eqs. (\ref{SumRuleNonPhysical})  and (\ref{RoughRelationIB}) in terms of the physical branching ratios and $CP$ asymmetries with $\pi^0$s containing in the final states.
The $CP$-asymmetry sum rule and the $CP$-asymmetry difference now read
\begin{equation}\label{SumRulePhysical}
A_{CP}^{K^+\pi^-}+\frac{\mathcal{B}(K^0\pi^+)}{\mathcal{B}(K^+\pi^-)}\frac{\tau_0}{\tau_+}A_{CP}^{K^0\pi^+} =\frac{2\mathcal{B}(K^+\pi^0)}{\mathcal{B}(K^+\pi^-)}\frac{\tau_0}{\tau_+}A_{CP}^{K^+\pi^0}+\frac{2\mathcal{B}(K^0\pi^0)}{\mathcal{B}(K^+\pi^-)}\frac{\tau_0}{\tau_+}A_{CP}^{K^0\pi^0}
-\Delta_{\text{IB}},
\end{equation}
and
\begin{equation}\label{RoughRelationIBPhysical}
A_{CP}^{K^+\pi^0}-A_{CP}^{K^+\pi^-}\approx\Delta_{\text{IB}},
\end{equation}
respectively, where $\Delta_{\text{IB}}$ accommodates the  $\pi^0\!\!-\!\eta\!-\!\eta'$ mixing correction and takes the form
\footnote{Contributions from the branching ratios, which are proportional to $ A_{CP}^{K^{+,0}\pi^0} \epsilon^{(\prime)}$, have been neglected in $\Delta_{\text{IB}}$.}
\begin{eqnarray}
&&\Delta_{\text{IB}}=\frac{2\mathcal{B}(K^+\pi^0)}{\mathcal{B}(K^+\pi^-)}\frac{\tau_0}{\tau_+}\Delta_{\text{IB}}^{K^+\pi^0 } +\frac{2\mathcal{B}(K^0\pi^0)}{\mathcal{B}(K^+\pi^-)}\frac{\tau_0}{\tau_+}\Delta_{\text{IB}}^{ K^0\pi^0}.
\end{eqnarray}
One interesting behaviour of this modification is that although Eq. (\ref{RoughRelation}) relates only the $CP$ asymmetries of $B^0\to K^+\pi^-$ and $B^+\to K^+\pi^0$, the isospin-breaking correction term $\Delta_{\text{IB}}$ in Eq. (\ref{RoughRelationIBPhysical}), however, not only contains the contribution of the process $B^+\to K^+\pi^0$, but also contains that of the process $B^0\to K^0\pi^0$.
The latter turns out to be numerically even more important in $\Delta_{\text{IB}}$.

We are now in a position to estimate the impact of the the isospin-breaking correction term $\Delta_{\text{IB}}$ to the $K\pi$-puzzle.
Although the amplitudes in Eqs. (\ref{DeltaIBKpm}) and (\ref{DeltaIBK0}) can be calculated theoretically or extracted from data, the four relative strong phases between these amplitudes, $\theta$, $\theta'$, $\tilde{\theta}$, and $\tilde{\theta}'$, are non-perturbative, preventing us from an accurate prediction of $\Delta_{\text{IB}}$.
What we can do is to give an rough estimation of the possible range of $\Delta_{\text{IB}}$ by treating the four strong phases as free parameters varying from 0 to $2\pi$ independently.
Based on this strategy, with the amplitudes borrowed from Ref. \cite{Cheng:2014rfa}, $\Delta_{\text{IB}}$ is estimated to be
\begin{equation}\label{eq:rangeDeltaIB}
  \Delta_{\text{IB}}=(-0.37,+0.37)\times\epsilon+(-1.16,+1.16)\times\epsilon'.
\end{equation}

In TABLE \ref{tab:rangDeltaIB}, with different values of $\epsilon$ and $\epsilon'$ from different references, the corresponding allowed range of $\Delta_{\text{IB}}$ is calculated via Eq. (\ref{eq:rangeDeltaIB}).
Note that the mixing parameters $\epsilon$ and $\epsilon'$ take quite different values throughout the literature, resulting in quite different ranges of the $\Delta_{\text{IB}}$, as is presented in the last column of this table.
From TABLE \ref{tab:rangDeltaIB} one can see that the isospin-breaking-correction term $\Delta_{\text{IB}}$ can be as large as a few percent, which indicates that the influence of the $\pi^0\!\!-\!\eta\!-\!\eta'$ mixing effect to the $CP$-asymmetry sum rule of $B\to K\pi$ and the $CP$-asymmetry difference of the $B^+\to K^+\pi^0$ and $B^0\to K^+\pi^-$ can not be simply ignored.
It is even possible that the $K\pi$-puzzle can be explained by the $\pi^0\!\!-\!\eta\!-\!\eta'$ mixing effect {\it alone}.
Take the $CP$-asymmetry difference between $B^+\to K^+\pi^0$ and $B^0\to K^+\pi^-$ as example.
If we treat the $\Delta_{\text{IB}}$-term as an uncertainty, with the mixing parameters from Ref. \cite{Escribano:2020jdy}, the $CP$-asymmetry difference now should read as
\begin{equation}
A_{CP}^{K^+\pi_3}-A_{CP}^{K^+\pi^-}=0.115\pm0.014\pm0.041_{\pi^0\!\!-\!\eta\!-\!\eta'}=0.115\pm0.043,
\end{equation}
from which one can clearly see that the uncertainty caused by the $\pi^0\!\!-\!\eta\!-\!\eta'$ mixing effect can be even larger than the other experimental uncertainties combined.
Hence the significance of the nonzeroness of this $CP$ difference is considerably reduced to less that 3 standard deviations.

\begin{table}[h!]
  \begin{center}
    \caption{The $\pi^0\!\!-\!\eta\!-\!\eta'$ mixing parameters quoted from different references and the corresponding range of $\Delta_{\text{IB}}$. Only the central values of $\epsilon$ and $\epsilon'$ are used when obtaining the range of $\Delta_{\text{IB}}$.}
    \label{tab:rangDeltaIB}
\begin{tabular}{c|c|c|c}
  \hline
  reference & $\epsilon (\%)$ & $\epsilon' (\%)$ & $\Delta_{\text{IB}} (\%$) \\
  \hline
  Kroll \cite{Kroll:2005sd} & $1.7\pm0.2$ & $0.4\pm0.1$ & $(-1.1,+1.1)$ \\
  Escribano \& Royo\cite{Escribano:2020jdy} & $0.1\pm0.9$ & $3.5\pm0.9$ & $(-4.1,+4.1)$ \\
  Benayoun et. al. \cite{Benayoun:2021ody} & $4.16\pm0.20$ & $1.05\pm0.05$ & $(-2.8,+2.8)$ \\
  \hline
\end{tabular}
  \end{center}
\end{table}

\section{Conclusion}
In conclusion, the contribution of the isospin-breaking $\pi^0\!\!-\!\eta\!-\!\eta'$ mixing effect to the $CP$ asymmetries of $B\to K\pi$ is investigated for the first time in this paper.
It is found that the $\pi^0\!\!-\!\eta\!-\!\eta'$ mixing effect pollutes the $K\pi$-puzzle more dramatically than the naive expectation.
This pollution is imbedded in the parameter $\Delta_{\text{IB}}$, which brings quite a large uncertainty to the $CP$-asymmetry sum rule of $B\to K\pi$ and the $CP$-asymmetry difference of the $B^+\to K^+\pi^0$ and $B^0\to K^+\pi^-$.
The analysis of this paper shows that it is even possible to explain the $K\pi$-puzzle by the $\pi^0\!\!-\!\eta\!-\!\eta'$ mixing effect {\it alone}.
Consequently, the implications of the $K\pi$-puzzle should be reconsidered.

\begin{acknowledgments}
This work was supported by National Natural Science Foundation of China under Contracts Nos. 11705081 and 12192261.
\end{acknowledgments}

\bibliography{zzhbib}
\end{document}